\documentstyle[11pt,epsf,epsfig]{article}
\begin{document}
\baselineskip 0.6cm
\renewcommand{\theequation}{\thesection.\arabic{equation}}
\newcommand{\gsim}{ \mathop{}_{\textstyle \sim}^{\textstyle >} }
\newcommand{\lsim}{ \mathop{}_{\textstyle \sim}^{\textstyle <} }
\newcommand{\vev}[1]{ \langle {#1} \rangle }
\newcommand{\EV}{ {\rm eV} }
\newcommand{\KEV}{ {\rm keV} }
\newcommand{\MEV}{ {\rm MeV} }
\newcommand{\GEV}{ {\rm GeV} }
\newcommand{\TEV}{ {\rm TeV} }
\newcommand{\barr}[1]{ \overline{{#1}} }
\newcommand{\del}{\partial}
\newcommand{\nn}{\nonumber}
\newcommand{\ra}{\rightarrow}
\newcommand{\bino}{\tilde{\chi}}
\def\tr{\mathop{\rm tr}\nolimits}
\def\Tr{\mathop{\rm Tr}\nolimits}
\def\Re{\mathop{\rm Re}\nolimits}
\def\Im{\mathop{\rm Im}\nolimits}
\setcounter{footnote}{1}

\begin{titlepage}

\begin{flushright}
UT-02-06\\
\end{flushright}

\vskip 3cm
\begin{center}
{\Large \bf Investigation of noscale supersymmetry breaking models
 \\ with a gauged U(1)$_{B-L}$ symmetry}
\vskip 2.4cm
Masaaki Fujii and Koshiro Suzuki

\vskip 0.4cm

{\it Department of Physics, University of Tokyo, 
 Tokyo 113-0033, Japan}

\vskip 4cm

\abstract{Noscale supersymmetry (SUSY) breaking model is investigated in the minimal
extension of the minimal supersymmetric standard model (MSSM) with a
gauged U(1)$_{B-L}$ symmetry.
We specifically consider a unification-inspired model
with the gauge groups
SU(3)$_{C}\times$SU(2)$_{L}\times$U(1)$_Y$ $\times$U(1)$_{B-L}$ 
$\subset$SU(5)$\times$ U(1)$_{5}$ for illustration.
While the noscale boundary condition at the grand unification 
scale ($M_{G}\simeq 2\times 10^{16}\GEV$) in the MSSM is not consistent with
phenomenological constraints, we show that it is if the gaugino of the
U(1)$_5$ multiplet is several times heavier than the gauginos of the
MSSM. However, if SU(5)$\times$U(1)$_5$ is further embedded in a larger
simple group, e.g. SO(10), the noscale boundary condition at $M_{G}$ is
inconsistent with phenomenological constraints. 
If we relax the noscale boundary condition and allow
non-zero soft scalar masses for the Higgs fields which spontaneously
break the U(1)$_{5}$ symmetry, 
the resultant spectrum of SUSY particles becomes consistent with all
the phenomenological constraints, even if we impose the GUT relation 
on the gauge coupling and the gaugino mass of the U(1)$_{5}$.
In this case,
the SUSY CP problem is also solved, since the condition $B\mu =0$ at the
boundary  can be imposed consistently with the electroweak 
symmetry breaking.
}

\end{center}
\end{titlepage}

\renewcommand{\thefootnote}{\arabic{footnote}}
\setcounter{footnote}{0}

%
%
%
%

\section{Introduction}
Low energy supersymmetry (SUSY) is expected to serve as a basis for
physics beyond the Standard Model (SM). If we just consider the minimal
supersymmetric standard model (MSSM) with generic soft SUSY breaking
terms as an effective low energy theory, then we must face with over a
hundred of additional parameters. However, although the SUSY particles
are not discovered at this moment, we already know from some low energy
experiments, such as detecting flavor changing neutral currents (FCNC)
and CP violation~\cite{FCNCandCP}, that these parameters cannot be
generic. These experiments give us a hint what kind of pattern the soft
SUSY breaking terms should have.

In this paper, we concentrate on models which have the so called
``noscale boundary condition''~\cite{noscale}.  In such a model, all
soft breaking terms except the gaugino masses are assumed to vanish at
some high energy scale $M_{X}$, which is usually taken as the GUT scale
$M_{G}\simeq 2\times 10^{16}\GEV$.  Soft breaking terms except gaugino
masses at the weak scale are generated by renormalization group effects
dominated by the gaugino loops, which are automatically flavor blind and
naturally suppress the FCNC interactions~\cite{IKYY}.  Furthermore, if
we can set the $B\mu$ term to be zero at the boundary scale $M_{X}$,
consistently with the radiative electroweak symmetry breaking, the SUSY CP
violation problem is also solved.  Therefore, under the assumption of
the noscale boundary condition, we can naturally avoid the SUSY FCNC and
CP violation problems and get a phenomenologically desirable, highly
predictive mass spectrum of SUSY particles. Recently, models with the
noscale boundary condition begin to attract much attention, since a
natural and a simple geometrical realization was proposed, i.e. the
gaugino-mediated SUSY breaking models~\cite{gauginomsb}.

However, 
it was shown recently that the minimal noscale model with $M_X = M_{G}$
is actually not consistent with phenomenological bounds \cite{ENO,
KomineYamaguchiNS, Komine}, mainly due to the lower bound on the Higgs
boson mass and the cosmological requirement that a charged particle (in
particular stau) is not the lightest SUSY particle (LSP). 
There might be several ways to
reconcile the noscale boundary condition with these phenomenological bounds,
e.g. by imposing non-universal gaugino masses \cite{KomineYamaguchiNS},
or by imposing the noscale boundary condition above the GUT scale
\cite{aboveGUT,Komine}.

In this paper, we propose another way, which is to change the mass
spectrum of SUSY particles by gauging some symmetry. We consider the
minimal extension of the MSSM by adding a gauged U(1)$_{B-L}$ symmetry
to the MSSM gauge groups.  Actually, the U(1)$_{B-L}$ symmetry is the
unique global symmetry which can be gauged without introducing 
any particles charged under the MSSM gauge groups. 
Furthermore, the existence
of three right-handed Majorana neutrinos is automatically required by
the anomaly cancellation condition of this symmetry. They naturally get
large masses of the order of the $B-L$ breaking scale, which allows
us to have a realistic mass spectrum of the lighter neutrinos via the
``seesaw'' mechanism~\cite{seesaw}.  Gauging the U(1)$_{B-L}$ symmetry is also
motivated from obtaining an exact R parity \cite{R-parity}.

In this work, we consider the minimal extension of the 
MSSM  with a gauged U(1)$_{B-L}$ symmetry and
analyze whether the noscale boundary condition at the GUT scale
is consistent with phenomenological bounds or not.

The organization of this paper is as follows. In section 2, we explain
the setup of our model.  We concentrate on the SU(5)$\times$U(1)$_{5}$
unification-inspired model. Here the U(1)$_{5}$ is the so called
``fiveness'', which is a linear combination of the weak hypercharge
U(1)$_{Y}$ and the U(1)$_{B-L}$, and we assume that this U(1)$_{5}$
symmetry is spontaneously broken at an intermediate scale. We also
discuss the subtlety of the mixing between the U(1)$_{Y}$ and the
U(1)$_{5}$, which arises due to the decoupling of the colored Higgs
fields.  In section 3, we show the results of the analyses,
and compare the differences between noscale models with and without a
gauged U(1)$_{B-L}$ symmetry. In section 4, we consider some variations
which relax the noscale boundary condition, and analyze whether they are
consistent with phenomenological constraints. 
Section 5 contains the summary and
concluding remarks.

\section{Models with a gauged $U(1)_{B-L}$ symmetry}
\setcounter{equation}{0}
We consider the minimal extension of the MSSM with gauge groups
SU(3)$_{C}\times$SU(2)$_{L}\times$U(1)$_{Y}\times$U(1)$_{5}$ $\subset$
SU(5)$\times$U(1)$_5$ below the GUT scale. The superpotential is given
by the following simplest form,
\begin{equation}
W = (y_{\nu})_{ij}\bar{N}_{i}L_{j}H_{u}+\lambda_{1}X(S\bar{S}-v^2)+
\frac{1}{2}(\lambda_2)_{ij}S\bar{N}_{i}\bar{N}_{j}+W_{\rm{MSSM}}, \nn \\
\label{superpotential}
\end{equation}
where $\lambda_1, \lambda_2, \;y_{\nu}$ are dimensionless coupling
constants, $v$ is the vacuum expectation value for the 
$S$ and $\bar{S}$ fields, which roughly corresponds to the $B-L$ breaking scale, and $W_{\rm MSSM}$ is the
superpotential of the MSSM,
\begin{equation}
 W_{\rm MSSM} = (y_u)_{ij} \bar{u}_{i}Q_{j}H_{u} 
- (y_d)_{ij}\bar{d}_{i}Q_{j}H_{d} - (y_e)_{ij}\bar{e}_{i}L_{j}H_{d}
+ \mu H_u H_d.
\end{equation}
Here, $\bar{N}_{i}$ is the chiral superfield of the right-handed Majorana
neutrino, and $X,\;S,\;\bar{S}$ are those which are responsible for the
$B-L$ symmetry breaking.  All of these extra chiral superfields are
singlets under the MSSM gauge groups.  The complete list of the matter
content of our model and the U(1) charge assignment are given in Table
1.
\begin{table}[h!]
\label{tab:charge}
\begin{center}
\begin{tabular}[t]{|c|c c c c|c c c c c c c|} \hline
field&$X$&$\bar{N}_{i}$&$S$&$\bar{S}$&$Q_{i}$&$\bar{u}_{i}$&$\bar{e}_{i}$&$\bar{d}_{i}$&$L_{i}$
 &$H_{u}$&$H_{d}$\\ \hline
$Q_{5}\times2\sqrt{10}$ & $0$ & $-5$ & $10$ & $-10$ & $-1$ & $-1$ & $-1$ &
 $3$ & $3$ & $2$ & $-2$\\ \hline
$B-L$& $0$ & $1$ & $-2$ & $2$ & $1/3$ & $-1/3$& $1$ &$-1/3$ & $-1$ & $0$
 &$0$ \\ \hline  
\end{tabular}
\end{center}
\caption{The list of the matter content and the U(1) charge
 assignment of our model. Here, the subscript ``$i$'' denotes the
 generation and runs 1,2,3. The charge of the U(1)$_{5}$, $Q_{5}$, is given
 by  the normalization consistent with the unification into E$_{6}$ and
 any of its subgroups.}
\end{table}
In that table, the charge $Q_{5}$ of the U(1)$_{5}$ is given by the 
normalization consistent with the unification into E$_{6}$ and any
of its subgroups. The U(1)$_{B-L}$ charge $B-L$ is 
given by a linear combination of the weak hyper charge $Y$ and $Q_{5}$
as 
\begin{equation}
B-L=-\frac{1}{5} (2\sqrt{10}\;Q_{5})+\frac{4}{5}Y.
\end{equation}

Now, we are at the point to discuss the renormalization group equations (RGEs).
There are some subtleties caused by  the kinetic term mixing between the two
U(1) gauge multiplets.  This is because ${\rm{Tr}}[Y Q_{5}]\neq 0$
below the GUT scale due to the decoupling of colored Higgs fields. 
(Here, ${\rm{Tr}}$ is taken with all the chiral superfields.)
After we perform a rotation on the two U(1) gauge multiplets 
to diagonalize their kinetic terms, there appear mixings in the
couplings between the matter fields $\phi_{i}$ and the two U(1) gauge
fields. We parameterize these couplings as follows:
\begin{eqnarray}
D_{\mu}\phi_{i}&=&\left(
\partial_{\mu}+i\left[\bar{g}_{Y}^{i}A_{\mu}^{Y}+\bar{g}_{5}^{i}A_{\mu}^{5}
\right]\right)\phi_{i}\;,\nonumber\\
\bar{g}_{Y}^{i}&=&g_{Y}Y^{i}+g_{5,Y}Q_{5}^{i}\;,\nonumber\\
\bar{g}_{5}^{i}&=&g_{Y,5}Y^{i}+g_{5}Q_{5}^{i}\;.
\label{couplings}
\end{eqnarray}
Here, the index ``$\: i\:$'' runs through all the chiral superfields.
Note that, in the following discussion, we choose the GUT normalization
also for the weak hypercharge; for example, $Y^{Q}=\sqrt{3/5}\;(1/6)$.

The one-loop RGEs of the gauge couplings 
can be obtained by following the methods given in Ref.~\cite{U(1)mixing} as:
\begin{eqnarray}
\frac{d}{dt}\left(
\begin{array}{cc}
g_{Y} & g_{Y,5}\\
g_{5,Y} & g_{5}
\end{array}
\right)= \frac{1}{16 \pi^{2}}\left(
\begin{array}{cc}
g_{Y}&g_{Y,5}\\
g_{5,Y}& g_{5}
\end{array}
\right)
\times\left(
\begin{array}{cc}
b_{Y}&b_{5,Y}\\
b_{Y,5}&b_{5}
\end{array}
\right)\;,
\label{RGEforgauge}
\end{eqnarray}
where 
\begin{eqnarray}
b_{Y}&=&{\rm{Tr}}\left[\bar{g}_{Y}\bar{g}_{Y}\right]=
\frac{33}{5}g_{Y}^{2}+\frac{57}{5}g_{5,Y}^{2}+
\frac{2\sqrt{6}}{5}g_{Y}g_{5,Y}\;,\nonumber\\
b_{5}&=&{\rm{Tr}}\left[\bar{g}_{5}\bar{g}_{5}\right]=
\frac{33}{5}g_{Y,5}^{2}+\frac{57}{5}g_{5}^{2}+\frac{2\sqrt{6}}{5}
g_{Y,5}g_{5}\;,\nonumber\\
b_{5,Y}&=&b_{Y,5}={\rm{Tr}}\left[\bar{g}_{5}\bar{g}_{Y}\right]=
\frac{33}{5}g_{Y}g_{Y,5}+\frac{\sqrt{6}}{5}\left(
g_{Y}g_{5}+g_{5,Y}g_{Y,5}\right)+\frac{57}{5}g_{Y}g_{5,Y}\; ,
\label{bfactor}
\end{eqnarray}
and $t\equiv{\rm{ln}}(\mu/\mu_{0})$.

Note that, because of the kinetic term mixing, $g_{Y}$ does not
follow the same RGE as in the MSSM. Even if we set the off diagonal
gauge couplings to be zero at some scale, they develop nonzero values
because of the non-vanishing ${\rm{Tr}}[YQ_{5}]$.  Then, how can we
discuss the gauge coupling unification?  Actually, the gauge field
$A_{\mu}^{Y}$ also couples with the fields which have nonzero U(1)$_5$
charges. Hence, the basis given in Eq.~(\ref{couplings}) is inadequate
to discuss the running of the unbroken U(1)$_{Y}$ gauge coupling
$g_{1}$.  We move to the on-shell basis in which one of the U(1) gauge
fields couples with only the fields which have zero U(1)$_5$ charges by
rotating the gauge multiplets. We can extract the unbroken U(1)$_Y$
gauge coupling $g_{1}$ from interactions of matter fields with this
massless gauge field as
\begin{equation}
g_{1}=\frac{g_{Y}g_{5}-g_{Y,5}g_{5,Y}}{\sqrt{g_{5}^2+g_{5,Y}^2}}\;.
\label{SMgY}
\end{equation}
Using Eqs.~(\ref{RGEforgauge}) and (\ref{bfactor}), one finds that 
$g_{1}$ defined by Eq.~(\ref{SMgY}) satisfies the one-loop RGE
\begin{equation}
\frac{d}{dt}g_{1}=\frac{1}{16\pi^2}\frac{33}{5}g_{1}^3 \: ,
\end{equation} 
which is the same as in the MSSM, both above and below the $B-L$
breaking scale $v$. Therefore, the condition for the unification of $g_{1}$
with SU(3)$_{C}$ and SU(2)$_{L}$ gauge couplings $g_{3}$ and $g_{2}$ is unaffected
by mixing, up to two-loop and threshold effects. Thus, the gauge coupling
unification is not spoiled in this model, and it is sensible to impose
the gauge coupling unification at the GUT scale:
\begin{equation}
g_{3}=g_{2}=g_{1}\equiv g_{U},\;\;\;g_{Y,5}=g_{5,Y}=0\;.
\label{g-unif}
\end{equation}
Note that this condition indicates $g_{Y}=g_{1}$, while $g_5$ is still
undetermined at the boundary.
We will assume this boundary condition throughout this
paper.

The RGEs for the gaugino masses are given by
\begin{eqnarray}
\frac{d}{dt}\left(
\begin{array}{cc}
M_{Y}& M_{Y,5}\\
M_{5,Y}& M_{5}
\end{array}\right)=
\frac{2}{16 \pi^2}\left(
\begin{array}{cc}
M_{Y}&M_{Y,5}\\
M_{5,Y} & M_{5}\end{array}
\right)\times\left(
\begin{array}{cc}
b_{Y}&b_{5,Y}\\
b_{Y,5}&b_{5}
\end{array}\right)\;,
\label{gauginomass}
\end{eqnarray}
where $b_Y, b_{5,Y}, b_{Y,5}, b_5$ are identical to those in
Eq.(\ref{bfactor}). 
These mass terms of the two U(1) gauginos
and their interactions between the matter fields are given by
\begin{eqnarray}
V&\supset& \sum_{i}\left[
\sqrt{2}(\bar{g}_{Y}^{i}\phi_{i}^{*}\lambda_{Y} \psi_{\phi_{i}}
+\bar{g_{5}}^{i}\phi_{i}^{*}\lambda_{5}\psi_{\phi_{i}})+
{\rm{h.c.}}
\right]\nonumber\\
&&\qquad\quad+\frac{1}{2}\left[\left(
\lambda_{Y},\lambda_{5}\right)
\left(
\begin{array}{cc}
M_{Y}& M_{Y,5}\\
M_{5,Y}& M_{5}
\end{array}\right)\left(
\begin{array}{c}
\lambda_{Y}\\\lambda_{5}
\end{array}
\right)+{\rm{h.c.}}
\right]\;,
\label{potential}
\end{eqnarray} 
where the index ``$\: i \:$'' runs through all the chiral superfields
as before, and $\psi_{\phi_{i}}$ is the fermion superpartners of
$\phi_{i}$.  $\lambda_{Y}$ and $\lambda_{5}$ are the gauginos which are
the superpartner of the gauge fields given in Eq.~(\ref{couplings}).
One can show that the well-known gaugino mass relation
\begin{equation}
\left(
\frac{M_{3}}{g_{3}^2},\;\frac{M_{2}}{g_{2}^2},\;
\frac{M_{1}}{g_{1}^2}\right)= \;{\rm const.} \: ,
\end{equation}
is precisely satisfied by the one-loop RGEs of this model.
Here, the mass of the gaugino which is the superpartner of the 
massless U(1)$_{Y}$ gauge field is given by
\begin{equation}
M_{1}=\frac{g_{5}^2 M_{Y}
-g_{5}g_{5,Y}(M_{5,Y}+M_{Y,5})+g_{5,Y}^2 M_{5}}{g_{5}^2+g_{5,Y}^2}\;,
\end{equation}
which is obtained by performing the same rotation on the gaugino fields
as the one performed on the two U(1) gauge fields. Thus, it is also
natural to impose the GUT relation among the gaugino masses with
vanishing off-diagonal elements,
\begin{equation}
M_{3}=M_{2}=M_{1}=M_{1/2},\;\;M_{Y,5}=M_{5,Y}=0 \:,
\label{M-unif}
\end{equation}
at the GUT scale. Note that, together with Eq.~(\ref{g-unif}), $M_Y$ is
equal to $M_1$, while $M_5$ is still undetermined at the boundary.
We will also assume
this boundary condition in the remaining part of this paper.

If we set the off-diagonal elements of the U(1) gauge couplings and
gaugino mass terms to be zero at the GUT scale, they remain fairly small
even at the intermediate scale,\footnote{If we assume the relations
given in Eq.~(\ref{ExGUT}), $\frac{
\sqrt{g_{5,Y}^2+g_{Y,5}^{2}}}{\sqrt{g_{Y}^{2}+g_{5}^{2}}} \sim 1\%$ and
$\frac{\sqrt{M_{5,Y}^{2}+M_{Y,5}^{2}}}{\sqrt{M_{Y}^{2}+M_{5}^{2}}} \sim
2\%$ at the intermediate scale $\mu\simeq 10^{10}\GEV$.
Even if we assume $g_{5}=5 g_{Y}$ and $M_{5}=5 M_{Y}$ at the GUT scale
$M_{G}$, these quantities only slightly increase to be $\sim 2\%,\;4\%$, respectively. 
} and hence many authors
neglect small effects caused by these mixings in their analyses of the
soft SUSY breaking parameters.  However, in our analyses, we take into
account the dominant mixing effects from the gauginos on the soft
scalar mass terms to determine them accurately.  The contribution of the
two U(1) gauginos to the RGE for soft scalar mass terms is calculated
as
\begin{eqnarray}
\frac{d}{dt}m_{i}^{2}\supset-\frac{1}{16\pi^2}\left[
\begin{array}{ccc}
8 \bar{g}_{Y}^{i2}|M_{Y}|^{2}+8\bar{g}_{5}^{i2}|M_{5}|^{2}+8(\bar{g}_{Y}^{i2}
+\bar{g}_{5}^{i2})|M_{\rm{off \: D}}|^{2}\\
\\
+8\bar{g}_{Y}^{i}\bar{g}_{5}^{i}\left(
M_{\rm{off\: D}}(M_{Y}+M_{5})^{\dag}+{\rm{h.c.}}
\right)
\end{array}
\right]\;,
\label{RGEWM}
\end{eqnarray}
where $M_{\rm{off\: D}}\equiv (M_{Y,5}+M_{5,Y})/2$. In the following
analyses, when we assume that $g_{Y}\neq g_{5}$ and/or $M_{Y}\neq M_{5}$
at the GUT scale, we use Eq.~(\ref{RGEWM}) and neglect small effects of
the U(1) mixing on the other soft breaking terms, e.g. the SUSY
breaking trilinear terms (A terms).

On the other hand, a dramatic simplification occurs when we further
impose the GUT relation also on the U(1)$_{5}$ gauge coupling and gaugino
mass as
\begin{equation}
g_{Y}=g_{5}=g_{U},\qquad M_{Y}=M_{5}=M_{1/2}
\label{ExGUT}
\end{equation}
at the GUT scale with vanishing off-diagonal elements.
In this case, we can go to the basis where the two U(1) gauge couplings
and the corresponding two gaugino masses do not mix at {\it{
arbitrary scales below}} $M_{G}$.
The gauge interaction of a matter field $\phi_i$ is 
\begin{eqnarray}
D_{\mu}\phi_{i}\supset i(Y^{i},Q_{5}^{i})
\left(
\begin{array}{cc}
g_{Y}&0\\
0&g_{5}
\end{array}\right)
\left(\begin{array}{c}
A_{\mu}^{Y}\\
A_{\mu}^{5}
\end{array}\right)\phi_{i}
\end{eqnarray}
at $M_{G}$. The existence of the U(1) mixing can be seen from the traces
of the U(1) charges,
\begin{equation}
Q\equiv \left(
\begin{array}{cc}
{\rm{Tr}}[YY]&{\rm{Tr}}[YQ_{5}]\\
{\rm{Tr}}[Q_{5}Y]&{\rm{Tr}}[Q_{5}Q_{5}]
\end{array}\right)=\left(
\begin{array}{cc}
33/5& \sqrt{6}/5\\
\sqrt{6}/5&57/5
\end{array}\right)\;.
\end{equation}
We can go to the basis where $Q$ is diagonal by rotating the basis in
the following way,
\begin{eqnarray}
\begin{array}{c}
(g_{Y}^{\prime},g_{5}^{\prime})=(g_{Y},g_{5})R\;,\\
(A_{\mu}^{Y\prime},A_{\mu}^{5\prime})=(A_{\mu}^{Y},A_{\mu}^{5})R\;,\\
(\lambda_{Y}^{\prime},\lambda_{5}^{\prime})=(\lambda_{Y},\lambda_{5})R
\;,
\end{array}
\qquad
R^{-1}QR=\left(
\begin{array}{cc}
9+\sqrt{6}&0\\
0&9-\sqrt{6}
\end{array}\right)\;,
\end{eqnarray}
where $R$ is a $2\times 2$ rotational matrix.  In this basis, the
off-diagonal elements of the $\beta$ functions are zero,
$b_{Y^{\prime},5^{\prime}}=b_{5^{\prime},Y^{\prime}}=0$, and by virtue
of Eq.~(\ref{ExGUT}), the off-diagonal elements of the gauge couplings
and gaugino masses are also zero at the GUT scale.  Therefore, we need
not worry about the mixings of the U(1) gauge couplings and gaugino mass
terms at {\it any} scale, and hence the calculation of the RGEs can be
carried out straightforwardly. The price for the choice of this basis is
that the U(1) charges of the fields are now complicated.  The new
charges $Y^{\prime}$ and $Q_{5}^{\prime}$ are given by
\begin{eqnarray}
Y^{\prime}=\frac{\sqrt{3}-\sqrt{2}}{\sqrt{10}}Y+
\frac{\sqrt{3}+\sqrt{2}}{\sqrt{10}} Q_{5}\;,\nonumber\\
\nonumber\\
Q_{5}^{\prime}=-\frac{\sqrt{3}+\sqrt{2}}{\sqrt{10}}Y+\frac{\sqrt{3}-\sqrt{2}}{\sqrt{10}} Q_{5}
\label{diagonalbasis}
\end{eqnarray}
in this basis. In the remainder of this paper, we use this basis when we
assume the extended GUT relation given by Eq.~(\ref{ExGUT}).\footnote{
Actually, under the assumption of this relation, one can show that this
model is equivalent to the SO(10)-inspired SU(3)$_{C}\times$SU(2)$_L$
$\times$U(1)$_{R}\times$U(1)$_{B-L}$ model 
by performing a similar
rotation on the basis of the two U(1) charges.}
\section{Noscale boundary conditions in models with a 
gauged U(1)$_{B-L}$ symmetry}
\setcounter{equation}{0}

In this section we show the results of the analyses and their
implications. We work on the SU(5)$\times$U(1)$_5$ unification-inspired
model. The soft scalar masses are assumed to vanish at the GUT scale
$M_G$, $m_{0}=0$, and they are generated by the RG effects at lower
energies. We also assume the SUSY breaking trilinear terms $A_0$ to
vanish at $M_G$, but leave the SUSY breaking Higgs mass term ($B\mu$
term) to be generic for a while, since we do not have a reliable
explanation for the origin of the Higgsino mass term ($\mu$ term), and
hence also for the $B\mu$ term.  The condition for the radiative electroweak
symmetry breaking (EWSB) relates the absolute value of the $\mu$ term,
$|\mu|$, and $B\mu$ (both at the weak scale) with the $Z$ boson pole
mass $m_Z$ and the ratio of the two VEVs of the Higgs doublets, $\tan
\beta \equiv v_u / v_d$. We choose $\tan \beta$ to be a free parameter,
and then $|\mu|$ and $B\mu$ are predicted. We assume the three standard
model gaugino masses $M_1, M_2, M_3$ to be universal, but remain the
U(1)$_5$ gaugino mass $M_5$ to be free:
\begin{equation}
 M_3 = M_2 = M_1 = M_{1/2}, \;\; M_5 = {\rm free}.
\label{bcfiveness}
\end{equation}
Thus, the parameters of the model are the gaugino masses $M_{1/2}$ and
$M_5$, $\tan \beta$, and sgn($\mu$) (the sign of $\mu$). The set of the
parameters which corresponds to $B\mu=0$ at $M_G$ will be shown as a
hypersurface in the parameter space in the remaining analyses.  We
calculate the spectrum of the model and search for the parameter regions
which are consistent with all the phenomenological bounds.

\subsection{Analytical procedure and phenomenological bounds}

Let us see how our analyses are done.
We first evolve the boundary conditions at $M_G$, i.e.,
\begin{equation}
 \{m_{0}=A_{0}=0, \;\; M_1 = M_2 = M_3 = M_{1/2}, \;\; M_5 \},
\end{equation}
down to the U(1)$_{B-L}$ breaking scale $g_{5} v$ by the one-loop RGEs. In
addition to the MSSM parameters, these include the additional U(1)
gauge coupling, the soft mass terms for the fields $S, \barr{S}, X,
\barr{N}_i$, the Yukawa couplings $y_\nu, \lambda_1, \lambda_2$, and the
corresponding $A$ terms (see Eq.(\ref{superpotential})). 
The one-loop RGEs are presented in the Appendix.

In our analyses, however, we neglect the Dirac neutrino Yukawa coupling
$y_{\nu}$ and the effects induced by this coupling. This is
justified if we use a relatively low $B-L$ breaking scale, which allows
us to obtain a conservative bound for the slepton masses. Small
$y_{\nu}$ also allows us to neglect the threshold effects at the $B-L$
breaking scale.  Actually this assumption is preferable to avoid the large
lepton-flavor-violating-interaction (LFVI) rates.  The off-diagonal elements in
the slepton masses at the scale $\mu$ are roughly given by
\begin{equation}
(m^{2}_{\tilde{L}})_{ij}\sim \frac{1}{16\pi^{2}} m_{0}^{2}(y_{\nu}^{\dag}y_{\nu})_{ij}{\rm{log}}
\left(\frac{M_{G}}{\mu}\right)
\end{equation}
in the minimal supergravity (mSUGRA) scenario. In the models with
the noscale boundary condition, the universal soft scalar masses vanish at
the GUT scale, hence the rate of the LFVIs are
expected to be suppressed. However, in the presence of a gauged
U(1)$_{B-L}$ symmetry, there exists an additional contribution to the
slepton masses from the gaugino of this extra U(1) gauge multiplet. This
contribution is mainly induced at high energy scales near the GUT scale,
because of the non-asymptotic freedom of the U(1) gauge
symmetry. Therefore, in our model, it is expected that the rate of
the LFVIs are not so much suppressed compared with those of 
the mSUGRA scenario~\cite{LFV}.  In the following analyses, we set the $B-L$
breaking scale, which is roughly equal to the right-handed Majorana
neutrino masses, to be $10^{10}\GEV$. This is low enough to satisfy the
constraints coming from the LFVIs.
  
Secondly, we use the {\tt SOFTSUSY} code \cite{softsusy} to evolve the
gauge, Yukawa couplings and the soft SUSY breaking parameters down to
the weak scale.  The Dirac neutrino Yukawa coupling $y_\nu$ is not
included in this code, but this is not a problem; the effects of the
neutrino Yukawa coupling on the MSSM parameters disappear below the
$B-L$ breaking scale, at least at the one-loop level. The two loop
effects are expected to be negligibly small. 

This code does not only solve the RGEs, but calculates the one-loop self
energies of all the particles and determines the physical pole masses by
identifying the pole of the propagator \cite{BPMZ}. It determines the
physical mass spectrum which is consistent with the boundary conditions
at the high energy scale (the U(1)$_{B-L}$ breaking scale in this case)
and the low energy scale. The low energy boundary conditions are:
$\barr{{\rm{MS}}}$ masses of the quarks and leptons at energy scale
$Q=91.19$ GeV, top quark pole mass,
\footnote{The pole mass of the Higgs boson $m_h$ is quite sensitive to the top
quark pole mass $m_t$ (with its dependence $\del m_h / \del m_t >0$). We
set $m_t = 175.0$ GeV in our analyses.}
and $\barr{{\rm{MS}}}$ gauge couplings of SU(3) and U(1)$_{\rm em}$. This is
done by performing the following iteration procedure: (step1) evolve
boundary conditions at $M_G$ to the U(1)$_{B-L}$ breaking scale by
one-loop RGEs $\ra$ (step2) input the running ($\barr{\rm{DR}}$) soft
parameters to the {\tt SOFTSUSY} code $\ra$ (step3) {\tt SOFTSUSY} finds
a physical spectrum consistent with high and low energy boundary
conditions $\ra$ (step4) run the $\barr{\rm{DR}}$ gauge and Yukawa couplings
up to the U(1)$_{B-L}$ breaking scale by {\tt SOFTSUSY}, which are in
general different from those obtained at (step1) $\ra$ (step5) run the
$\barr{\rm{DR}}$ gauge and Yukawa couplings to high energies and determine
$M_G$, then run them back to the U(1)$_{B-L}$ breaking scale by one-loop
RGEs $\ra$ (step2) $\ra$ (step3) ... This iteration procedure is
performed until the gauge and Yukawa couplings match at the U(1)$_{B-L}$
breaking scale, that is, at (step2) and (step4). This procedure amounts
to taking into account the loop effects and the weak scale SUSY threshold
corrections, and it determines the mass spectrum precisely.

Determining the mass spectrum and the running $\barr{\rm{DR}}$
parameters (the gauge couplings, the Yukawa couplings, the $\mu$ term,
the $B\mu$ term, and the $A$ terms) allows us to identify the region of
the parameter space of the model which is excluded by particle search
experiments and cosmological requirements. We specifically consider the
lower bounds on the masses of the Higgs boson (114.1 GeV
\cite{Higgsmass}) \footnote{Actually, the lower bound on the Higgs boson
mass in the MSSM is somewhat weaker in large ${\rm{tan}}\beta$
region. However, in such a region, the most severe bound comes from the
BR($b\rightarrow s\gamma$) or the requirement of neutral LSP, and hence the allowed region
of the parameter space is not altered by adopting the Higgs boson mass
bound $(m_{h}=114.1\GEV)$. }and the selectron (99 GeV
\cite{selectronmass}) from the LEP experiments. We also consider the
ratio of the stau mass to the neutralino mass, $m_{\tilde{\tau}_1}/
m_{\tilde{\chi}}$, which should be larger than 1 to avoid charged
LSP. In addition, bounds from experiments of rare processes can also be
applied to restrict the parameter space. We consider the branching ratio
(BR) of $b \ra s \gamma$ in particular, since it provides a powerful
experimental testing ground for physics beyond the SM, because of its
sensitivity to virtual effects of new particles.  In this work, we
perform a calculation of 
${\rm{BR}}(b\rightarrow s\gamma)$ based on Ref.~\cite{bsg}, which
includes the dominant next-to-leading-order (NLO) corrections enhanced
by large ${\rm{tan}}\beta$ factors. For the conservative bound, we adopt
$2.0\times 10^{-4}<{\rm{BR}}(b\rightarrow
s\gamma) <4.5\times 10^{-4}$~\cite{bsgexp} from CLEO experiments.  The bounds we adopted cast the
most stringent constraints on the model.

We provide some comments on the Higgs boson mass.  As is explained in
Ref.\cite{softsusy}, the {\tt SOFTSUSY} code predicts the Higgs boson
mass to be systematically 2--4 GeV heavier than the combination of the
codes {\tt SSARD} \cite{ssard} and {\tt FeynHiggs} \cite{feynhiggs}. We
adopt {\tt SOFTSUSY}, since this code performs calculations of  full one-loop self energies or
accurate approximations of them based on Ref.~\cite{BPMZ} to determine the pole
masses of SUSY particles, and since it predicts the larger Higgs boson mass
which gives us a more conservative bound.
\subsection{Results and their implications}

\subsubsection{The minimal noscale model}

Before the results for the SU(5)$\times$U(1)$_5$ model, we show the
result for the minimal noscale model in
Figs.\ref{fig:msugraMup}. 
The figure on the left corresponds to the case $\mu>0$, and the one 
on the right corresponds to the case  $\mu<0$.
This model is the
conventional MSSM with the noscale boundary condition, i.e., $m_0 = 0$
for all scalar soft masses and nonzero universal gaugino masses $M_{1/2}
\neq 0$ at $M_G$. The red (solid) line is the contour of the Higgs boson
mass $m_h = 114.1$ GeV.
The region below this line is currently
excluded. The blue (dashed) line is the upper bound on $M_{1/2}$ from
the cosmological requirement that the stau is not the LSP. The green
(dotted) line is the lower bound on $M_{1/2}$ from the $b \ra s \gamma$
experiments. 
The orange (dot-dashed) line is the contour of the
right-handed selectron mass $m_{\tilde{e}_R} = 99$ GeV. 
The purple (solid) line in the case $\mu>0$ denotes the predicted 
${\rm{tan}}\beta$ when we set the condition $B\mu=0$ at 
the GUT scale.\footnote{As for the case $\mu<0$, the
RG effects of the gauginos on the $B\mu$ term 
is of the opposite sign
compared with the case $\mu>0$, and hence $B\mu=0$ at the GUT scale cannot be 
consistent with the EWSB.}

The black shaded region on
the upper right side is the part where the radiative electroweak symmetry
breaking (EWSB) cannot be implemented. The constraint from $b \ra s \gamma$
is much stringent in the case $\mu <0$, since the SUSY contribution to 
the BR$(b\rightarrow s\gamma)$ interfers constructively with the SM
contribution, which is opposite to the case $\mu>0$. 
However,
as is already shown in Refs.\cite{ENO, Komine}, we can see that there is
almost no region which simultaneously satisfies these constraints even
in the case $\mu >0$. The obstacle is mainly due to the fact that in
noscale models, the lightest neutralino (mostly bino) 
is nearly degenerate with the right-handed stau and,
unfortunately, is slightly heavier than the stau.

If we can alter the particle spectrum and make the stau heavier, the
requirement of neutral LSP becomes less restrictive and a wider
parameter region compatible with the constraints may arise. 
In this work, we work on the minimal extension of the MSSM with 
a gauged U(1)$_{B-L}$ symmetry. 
As we will see in the remaining sections, 
the extra positive contribution to 
the stau mass from the U(1)$_{B-L}$ gaugino loops makes the noscale
boundary condition consistent with all the constraints.
In addition, if we  relax the noscale boundary condition and
allow the non-zero soft scalar masses of the order of the 
gaugino mass for the fields $S,\;\bar{S}$ at the GUT scale,
the U(1)$_{B-L}$ $D$-term contribution gives us another 
solution for the charged LSP problem.

\subsubsection{SU(5)$\times$U(1)$_5$ model}

\begin{figure}[ht]
\begin{center}
\centerline{\psfig{figure=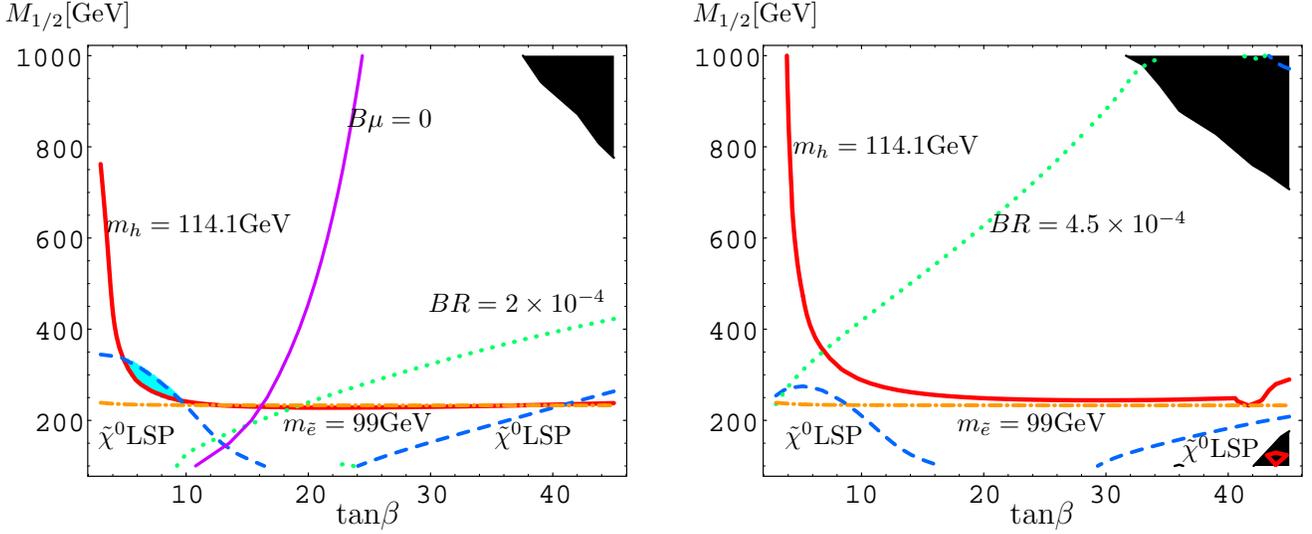,width=18cm}}
\end{center}
\begin{picture}(0,0)
 \put(23,150){{\bf{$m_{h}=114.1\GEV$}}}
 \put(145,120){{\bf{$BR=2\times 10^{-4}$}}}
 \put(114,190){{\bf{$B\mu=0$}}}
 \put(170,70){{\bf{$\tilde{\chi}^{0}$}}LSP}
 \put(20,70){{\bf{$\tilde{\chi}^{0}$}}LSP}
 \put(90,75){{\bf{$m_{\tilde{e}}=99\GEV$}}}
 \put(-15,230){{\bf{$M_{1/2}$}}[GeV]}
 \put(245,230){{\bf{$M_{1/2}$}}[GeV]}
\large 
\put(110,40){{\bf{${\rm{tan}}\beta$}}}  
\put(365,40){{\bf{${\rm{tan}}\beta$}}}
\small
  \put(283,180){{\bf{$m_{h}=114.1\GEV$}}}
 \put(357,150){{\bf{$BR=4.5\times 10^{-4}$}}}
 \put(430,65){{\bf{$\tilde{\chi}^{0}$}}LSP}
 \put(280,70){{\bf{$\tilde{\chi}^{0}$}}LSP}
 \put(345,75){{\bf{$m_{\tilde{e}}=99\GEV$}}}
\end{picture}
\vspace{-1.5cm}
\caption{Constraints on the ${\rm{tan}}\beta-M_{1/2}$ plane for the
 minimal noscale model with $\mu>0$ (left) and $\mu<0$ (right),
 respectively. The blue (dashed) line denotes the upper bound on
 $M_{1/2}$ for the neutralino to be the LSP. The purple (solid) line for
 the case $\mu>0$ denotes the parameter region predicted from the
 condition $B\mu=0$ at the GUT scale.  The other lines correspond to the
 lower bounds on $M_{1/2}$ from various constraints, which are those
 from the Higgs boson mass (red solid), the selectron mass (orange
 dot-dashed) and the BR($b\rightarrow s\gamma$) (green dotted),
 respectively.  In the black shaded region, the EWSB does not occur or
 tachyonic scalars emerge. The light shaded region is allowed (present
 only for the case $\mu>0$).  }  \label{fig:msugraMup}
\end{figure}

\begin{figure}[h!]
\begin{center}
\centerline{\psfig{figure=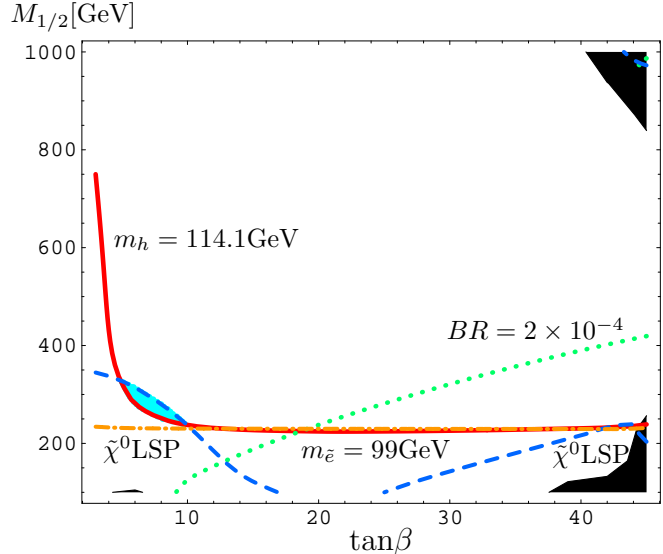,width=8.5cm}}
\begin{picture}(0,0)
\put(-86,120){{\bf{$m_{h}=114.1\GEV$}}}
\put(40,85){{\bf{$BR=2\times 10^{-4}$}}}
 \put(80,38){{\bf{$\tilde{\chi}^{0}$}}LSP}
 \put(-90,40){{\bf{$\tilde{\chi}^{0}$}}LSP}
 \put(-15,40){{\bf{$m_{\tilde{e}}=99\GEV$}}}
 \put(-125,205){{\bf{$M_{1/2}$}}[GeV]}
\large
\put(3,5){{\bf{${\rm{tan}}\beta$}}}
\small
\end{picture}
\end{center}\vspace{-1cm}
\caption{Constraints on the ${\rm{tan}}\beta-M_{1/2}$ plane for the
SU(5)$\times$U(1)$_{5}$ inspired model with $\mu>0$ and the extended GUT
relations given in Eqs.~(\ref{GUT1}) and (\ref{GUT2}). The conventions
 are the same as those in Figs.~\ref{fig:msugraMup}. The light shaded
 region is allowed.} 
\label{fig:LRmodel}
\end{figure}

Now we go to the SU(5)$\times$U(1)$_5$ model. First of all, consider the
case where the extended gauge coupling 
unification and the gaugino mass relation
are imposed:
\begin{equation}
 g_1 = g_2 = g_3 = g_5,
\label{GUT1}
\end{equation}
\begin{equation}
 \frac{M_1}{g_1^2} = \frac{M_2}{g_2^2} = \frac{M_3}{g_3^2} = \frac{M_5}{g_5^2}.\label{GUT2} 
\end{equation}
They are natural assumptions for the case in which the gauge groups
SU(5)$\times$U(1)$_{5}$ are embedded in a larger gauge
group, such as SO(10).\footnote{Here, we consider the case where the
SO(10) gauge group breaks down into the gauge groups
SU(3)$_{C}\times$SU(2)$_{L}$ $\times$U(1)$_{Y}\times$U(1)$_{5}$  in a
single step at the scale $M_{G}$.} The result for the 
case $\mu>0$ is shown in Fig.\ref{fig:LRmodel}. The conventions are the
same as those in the Figs.\ref{fig:msugraMup}.  The parameter region
consistent with $B\mu=0$ at the GUT scale is not shown in this figure,
since it is almost the same as in the minimal noscale model. The black
shaded regions in the lower and upper right side are the parameter
spaces where either the EWSB does not occur, or tachyonic scalars arise. We
can see that there is almost no region which is compatible with the  
constraints. The result for the case $\mu<0$ is not presented, but it is
also almost the same as in the minimal noscale model with $\mu<0$ shown
in the Figs.\ref{fig:msugraMup}.  To understand this result, we explain
the phenomenology of the SU(5)$\times$U(1)$_5$ model.

First of all, consider the ratio of the stau mass to the neutralino
mass, $m_{\tilde{\tau}_1}/ m_{\tilde{\chi}}$. The RGEs of the three
standard model gaugino masses are unaffected by gauging the U(1)$_{B-L}$
symmetry (as is explained in section 2). On the other hand, new sources
for the squark and slepton masses exist. One is the additional positive
RGE effects from the U(1)$_5$ gaugino mass, and another is the $D$-term
contributions due to the spontaneous breaking of the U(1)$_5$. The
$D$-term contribution to the mass squared of a scalar field $\phi_i$ is
approximately given by
\begin{equation}
 (\Delta m_i^2)_{D{\rm-term}} = \frac{1}{\sqrt{10}}(m_{\barr{S}}^2 -
m_{S}^2) Q_{5}^i, \label{Dterm}
\end{equation}
where $Q_{5}^i$ is the $U(1)_5$ charge of $\phi_i$ and $m_S^2,
m_{\barr{S}}^2$ are the soft masses for the fields $S, \barr{S}$.  Here,
small mixing effects are neglected.  In our numerical calculations, all
of these effects are included by using the diagonal basis given in
Eq.~(\ref{diagonalbasis}). This contribution is added at the $B-L$
breaking scale $g_{5}v$, and is renormalized down to lower energy
scales.  
As for the detailed discussions about the 
$D$-term contributions to soft scalar masses, see Refs.~\cite{Dterm1,Dterm2}.
Eq.(\ref{Dterm}) is zero at $M_G$ if we impose the noscale
boundary condition, but it is nonzero (and negative for particles with
negative $U(1)_5$ charges) at the breaking scale of the U(1)$_{5}$. 
This is because $m_S^2$ receives a negative
contribution from the renormalization by the Yukawa coupling $\lambda_2$
which is absent for $m_{\barr{S}}^2$, and hence $m_{\barr{S}}^2 - m_S^2
> 0$. Due to these effects, the mass spectrum is shifted from that of
the minimal noscale model.  In particular, the mass squared of the
right-handed slepton at the weak scale is approximately shifted by an
amount (neglecting the mixing effects)
\begin{eqnarray}
 \Delta m_{\tilde{e}_R}^2 &=&\frac{2(Q_{5}^{\tilde{e}_{R}})^{2}}{b_{5}} M_{5}^{2}
\left[
1-\frac{1}{\left[1+\frac{b_{5}}{2\pi}\alpha_{5}{\rm{log}}
\left(\frac{M_{G}}{\mu_{B-L}}\right)\right]^2}
\right]-\frac{1}{20}(m_{\barr{S}}^{2}-m_{S}^{2})\;,\nonumber\\
&\simeq& 0.0034 M_{5}^{2}-\frac{1}{20}(m_{\barr{S}}^{2}-m_{S}^{2}),
\label{sleptonmass}
\end{eqnarray}
where $M_{5}$ is the U(1)$_{5}$ gaugino mass at $M_G$ and
$m_{\barr{S}}^2,\;m_S^2$ are evaluated at the $B-L$ breaking scale. The
constants are defined as $b_{5}\equiv 57/5$ and
$Q_{5}^{\tilde{e}_{R}}(=-1/2\sqrt{10})$ is the U(1)$_{5}$ charge of the
right-handed selectron, and $\mu_{B-L}$ is the
$B-L$ breaking scale. In the second line, we use
$\alpha_{5}^{-1}(=\alpha_{\rm{GUT}}^{-1})=24$ and
$\mu_{B-L}=10^{10}\GEV$.\footnote{ After taking the mixing effects into
account, soft breaking masses increase by a small amount. For example,
if we assume the extended GUT relations given in Eqs.~(\ref{GUT1}) and
(\ref{GUT2}), the right-handed selectron mass squared
$m_{\tilde{e}_{R}}^2$ at the weak scale becomes about $1$\% larger by
including the mixing effects.}

This shows that the right-handed sleptons acquire positive soft masses
from the  U(1)$_5$ gaugino, and negative contributions from the U(1)$_5$
$D$ term. The $D$-term contribution depends on the size of the Yukawa
coupling $\lambda_{2}$. If we assume $\lambda_{2}$ has a comparable size
as the U(1)$_5$ gauge coupling $g_{5}$ at the U(1)$_{5}$ breaking scale,
this $D$-term contribution is relatively small.\footnote{ If we take
$\lambda_{2}\simeq 0.1g_{5}$ at the U(1)$_{5}$ breaking scale
($=10^{10}\GEV$), the D-term contribution is about $10\%$ of the gaugino
contribution, which is the first term of Eq.~(\ref{sleptonmass}). It
reaches $40\%$ of the gaugino contribution, if we take
$\lambda_{2}\simeq 0.8g_{5}$ at the breaking scale.  Through out this
paper, we assume $\lambda_{2}\simeq 0.1g_{5}$ at the U(1)$_{5}$ breaking
scale ($=10^{10}\GEV$) to obtain a conservative bound on the slepton
masses.  } Because of the small coefficients in Eq.~(\ref{sleptonmass}),
the resultant selectron mass is almost the same as in the minimal
noscale model.  The Higgs boson mass also remains the same as that in
the minimal noscale model since the contribution to the stop masses from
the U(1)$_{5}$ gaugino is also very small and the stop masses at the
weak scale are dominated by the gluino-loop contribution.  This is why
the lower bound on the Higgs boson mass and the requirement of neutral
LSP still conflict with each other.  Therefore, if we assume
that the gauge groups SU(5)$\times$U(1)$_{5}$ are embedded in a larger gauge group such as
SO(10), we cannot set the noscale boundary condition at the GUT scale
without additional assumptions.\footnote{Because of the reason mentioned
in footnote $2$, this is also true for the
SU(3)$_{C}\times$SU(2)$_{L}\times$U(1)$_{R}\times$U(1)$_{B-L}$ model
embedded in SO(10).}  If the SO(10) gauge group breaks down into the
gauge groups SU(5)$_{{\rm{GUT}}}\times$U(1)$_{5}$ above the GUT
scale, the conditions given in 
Eqs.~(\ref{GUT1}) and (\ref{GUT2}) do not hold.
However, the above conclusion is not altered, because the gauge 
coupling $g_{5}$ is smaller than those of the MSSM gauge groups at the
scale $M_{G}$
due to the non-asymptotic freedom of the U(1)$_{5}$.

\begin{figure}[ht]
\begin{center}
\centerline{\psfig{figure=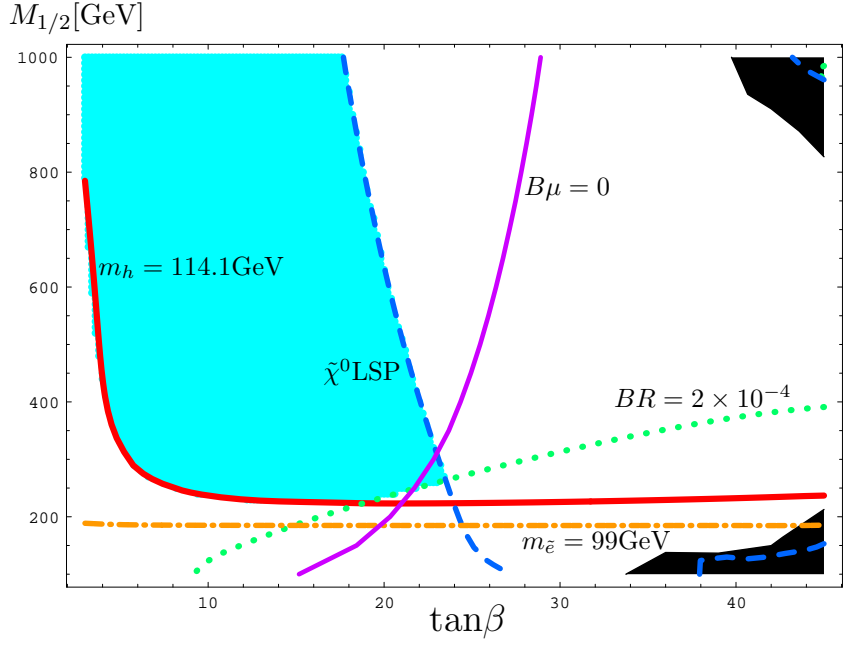,width=11cm}}
\begin{picture}(0,0)
\put(-125,140){{\bf{$m_{h}=114.1\GEV$}}}
\put(36,170){{\bf{$B\mu=0$}}}
\put(70,90){{\bf{$BR=2\times 10^{-4}$}}}
\put(35,36){{\bf{$m_{\tilde{e}}=99\GEV$}}}
\put(-40,100){{\bf{$\tilde{\chi}^{0}$}}LSP}
\normalsize
\put(-159,235){{\bf{$M_{1/2}$}}[GeV]}
\Large
\put(0,5){{\bf{$\rm{tan}\beta$}}}
\small
\end{picture}
\end{center}
\caption{The same as Fig.~\ref{fig:LRmodel} but with the condition 
$M_{5}=5 M_{1/2}$ at the GUT scale. Now there is a wide allowed region.}
\label{fig:Ftimes5}
\end{figure}

\begin{figure}[h!]
\begin{center}
\centerline{\psfig{figure=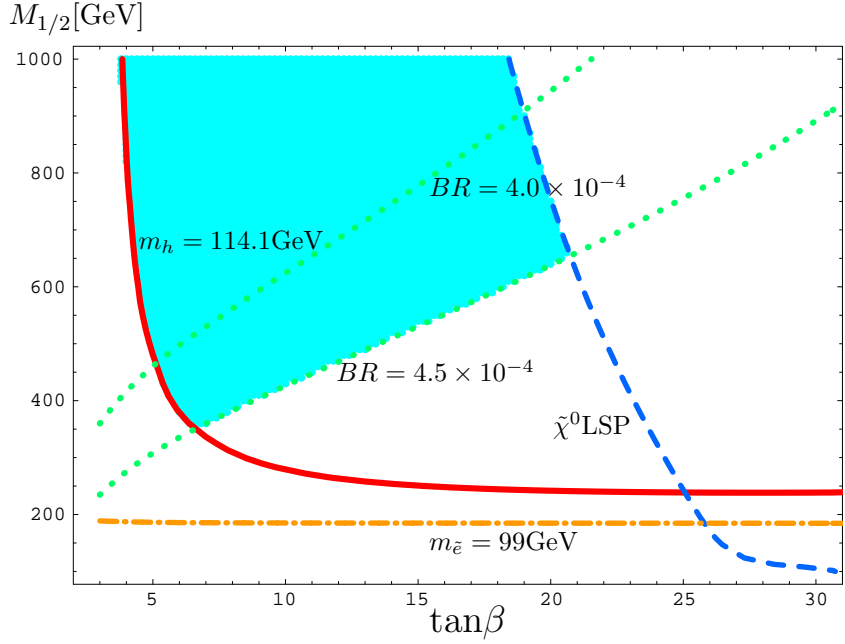,width=11cm}}
\begin{picture}(0,0)
\put(-110,150){{\bf{$m_{h}=114.1\GEV$}}}
\put(-35,100){{\bf{$BR=4.5\times 10^{-4}$}}}
\put(-0,170){{\bf{$BR=4.0\times 10^{-4}$}}}
\put(0,36){{\bf{$m_{\tilde{e}}=99\GEV$}}}
\put(47,80){{\bf{$\tilde{\chi}^{0}$}}LSP}
\normalsize
\put(-159,235){{\bf{$M_{1/2}$}}[GeV]}
\Large
\put(0,5){{\bf{$\rm{tan}\beta$}}}
\end{picture}
\end{center}
\caption{The same as Fig.~\ref{fig:Ftimes5} but for $\mu<0$.
Even in this case, a wide region is allowed.} 
\label{fig:Ftimes5Mun}
\end{figure}

Now, let us relax the extended GUT relations given in Eqs.~(\ref{GUT1})
and (\ref{GUT2}). In this case, we have no reason to expect that
the gauge coupling $g_{5}$ and the gaugino mass $M_{5}$ of the U(1)$_{5}$ are 
the same as those of the MSSM gauge groups, since the vector multiplet
of the U(1)$_{5}$ does not belong to the SU(5)$_{\rm{GUT}}$ vector
multiplet above the GUT scale.
In fact, such a SU(5) unification model (rather than a SO(10) GUT) is also desirable 
for obtaining bimaximal mixings among the lighter neutrinos\cite{LM,LMandPH},
since the leptons and quarks reside in the different multiplets.
In this case, we can easily obtain much larger slepton masses and 
satisfy all the phenomenological bounds by increasing $M_{5}$.
(One can easily see from Eq.~(\ref{sleptonmass}) that the resultant 
spectrum can be altered only slightly even if we 
significantly increase the gauge coupling $g_{5}$.) 
In the following representative examples of numerical calculations, we 
set the gaugino mass relation as
\begin{eqnarray}
&&M_{1}=M_{2}=M_{3}=M_{1/2}\;,\nonumber\\
&&\qquad\;\; M_{5}=5 \;M_{1/2}\;,
\label{LargeM5}
\end{eqnarray}
at the GUT scale. As for the relation between the gauge couplings, we
use Eq.~(\ref{GUT1}) because of the reason mentioned above.
  The results
for the cases  $\mu>0$ and $\mu<0$ are shown in Figs.\ref{fig:Ftimes5}
and \ref{fig:Ftimes5Mun}, respectively.
The conventions are the same as those in Figs.~\ref{fig:msugraMup}.
We can see that there exists a wide parameter region consistent with 
all the constraints, even for the case $\mu<0$.

\section{Relaxing the noscale boundary condition}
\setcounter{equation}{0}

So far we have considered a model with the noscale
boundary condition and a gauged U(1)$_{B-L}$ symmetry.
As we have seen, we can easily obtain a spectrum of SUSY particles
consistent 
with experimental and cosmological constraints by 
imposing the boundary conditions, e.g., Eqs.(\ref{GUT1}) and (\ref{LargeM5}).
These conditions are expected to be quite plausible 
in a SU(5) (not a SO(10)) unification model, which is preferable to 
explain the bimaximal mixings among the lighter neutrinos.
Unfortunately however, we can set the condition 
$B\mu=0$ at the GUT scale
only in a small restricted parameter region.\footnote{
This is true even if we assume a much larger mass for the gaugino
of the U(1)$_{5}$, $M_{5}$.} 
Therefore, in this model,
although the FCNC interactions are naturally suppressed by virtue of 
vanishing soft scalar masses at the GUT scale,
the SUSY CP problem requires an accidental  phase cancellation between 
the $B\mu$ term and the $\mu$ term, which 
are expected to have independent phases, in most of the parameter space.

However, there is a quite natural solution
to this problem by considering a variation of the model we are working on.
The gauged U(1)$_{B-L}$ symmetry inevitably
requires a set of new Higgs fields which are singlets under the MSSM
gauge groups to spontaneously break the $B-L$ symmetry at some high
energy scale.  They correspond to the fields $S,\;\bar{S}$ in our model.
By assigning even $B-L$ charges to these fields, as is done in our model,
there
is no allowed coupling to the MSSM fields at the
renormalizable level. (Exact R-parity conservation is also automatically
guaranteed.)
Therefore, there appears no FCNC problem 
even if we allow non-zero soft scalar masses for the fields
$S,\;\bar{S}$ to be of the 
order of the gaugino masses at the GUT
scale. 

These soft scalar masses arise from the nonrenormalizable couplings 
with the SUSY breaking fields,
\begin{equation}
 {\cal L} \supset \int d^4 \theta \; \left( \lambda_S \; \frac{Z^{\dag} Z}{M_{pl}^2} S^{\dag}S 
+ \lambda_{\barr{S}} \; \frac{Z^{\dag} Z}{M_{pl}^2} \barr{S}^{\dag} \barr{S} \right),
\end{equation}
where $Z$ stands for the SUSY breaking fields and $\lambda_S,
\lambda_{\barr{S}}$ are unknown dimensionless couplings. 
In this case, 
it is natural to expect that $m_S^2$ and $m_{\barr{S}}^2$
differ by a factor of order one, since there is no reason to believe
that the $\lambda_{S}$ and $\lambda_{\bar{S}}$ are degenerate.
Thus, the $D$-term contributions to
the soft scalar masses due to the breaking of U(1)$_{B-L}$ do not in
general vanish at tree level (see eq.(\ref{Dterm})). If $m_{S}^2 -
m_{\barr{S}}^2 >0$, the right-handed slepton masses acquire positive
$D$-term contributions (cf. $(2 \sqrt{10}\;Q_5^{\tilde{e}_R}) = -1$) while the
neutralino mass almost remains the same, and the requirement of neutral
LSP might be extremely relaxed. 
Note that this non-vanishing $D$ term also induces  relatively large 
RG effects through the tadpole diagram, which is proportional to 
$d m_{i}^{2}/dt\propto g_{5}^{2}Q_{5}^{i}S_{5}$ in the limit of vanishing
U(1) mixing effects. (See the Appendix for notations.)

Such a situation can be easily realized, for example, 
in gaugino-mediated SUSY 
breaking models by allowing the $S$ and $\bar{S}$ fields to
propagate in the bulk and to have contact interactions 
with SUSY breaking fields which reside in the hidden sector brane.

\begin{figure}[ht]
\begin{center}
\centerline{\psfig{figure=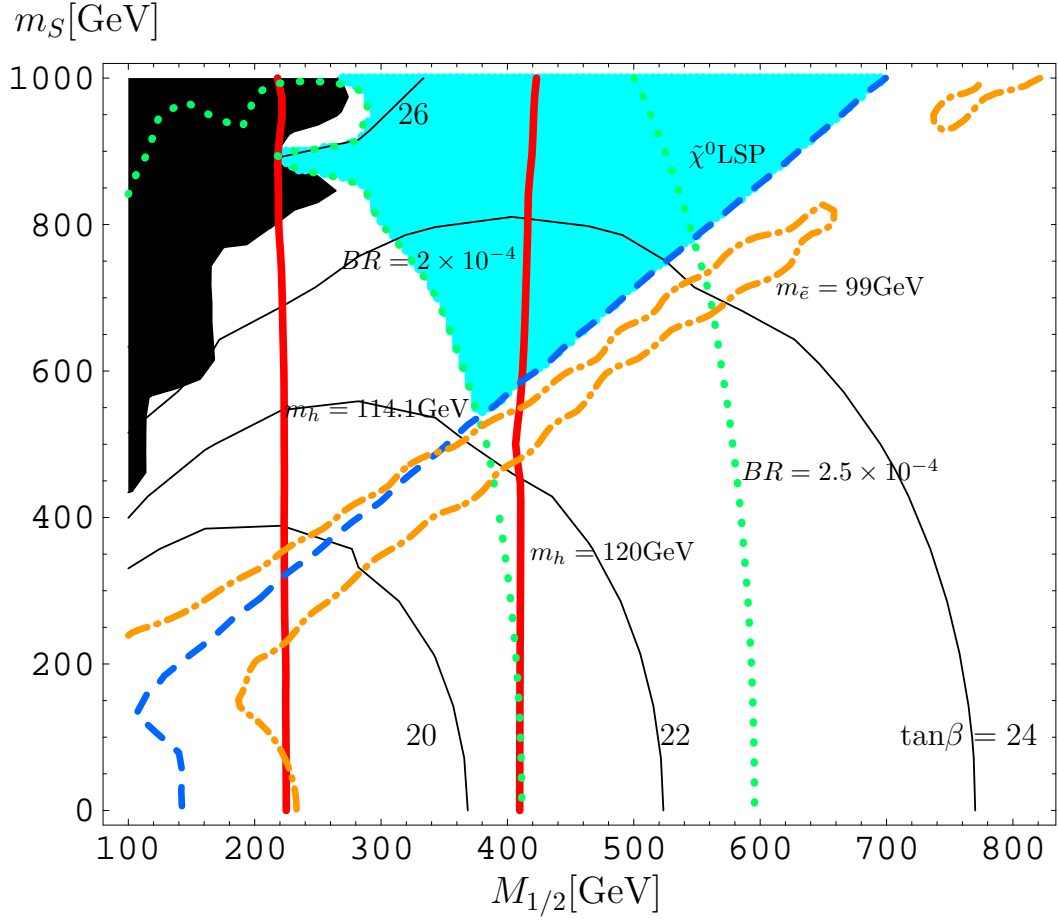,width=14cm}}
\begin{picture}(0,0)
\put(0,130){{\bf{$m_{h}=120\GEV$}}}
\put(-72,240){{\bf{$BR=2\times 10^{-4}$}}}
\put(60,280){{\bf{$\tilde{\chi}^{0}$}}LSP}
\put(80,160){{\bf{$BR=2.5\times 10^{-4}$}}}
\put(-93,184){{\bf{$m_{h}=114.1\GEV$}}}
\put(93,230){{\bf{$m_{\tilde{e}}=99\GEV$}}}
\large
\put(140,60){{\bf{$\rm{tan}\beta=24$}}}
\put(49,60){{\bf{$22$}}}
\put(-47,60){{\bf{$20$}}}
\put(-50,295){{\bf{$26$}}}
\Large
\put(-15,0){{\bf{$M_{1/2}$}}[GeV]}
\put(-195,330){{\bf{$m_{S}$}}[GeV]}
\end{picture}
\small
\end{center}
\caption{Constraints on the model with $m_{S}^{2}=-m_{\bar{S}}^{2}(\neq
0)$, $B\mu=0$ and with the conditions given in Eqs.~(\ref{GUT1}),
(\ref{GUT2}) at the GUT scale. The thin black (solid) lines are the
contours of the predicted ${\rm{tan}}\beta$, which are 20, 22, 24, and
26, respectively. Other conventions are the same as those in
Figs.~\ref{fig:msugraMup}.  } \label{fig:Dterm}
\end{figure}

In Fig.{\ref{fig:Dterm}}, we show the result when we vary $m_{S}^2$ and
$m_{\barr{S}}^2$. In this analysis, we take $m_S^2 = - m_{\barr{S}}^2$
at the GUT scale for simplicity, and assume the relations in
Eqs.~(\ref{GUT1}) and (\ref{GUT2}), i.e. the extended unification
conditions. We choose $\{ M_{1/2}, m_{S}^{2}\}$ as free parameters, and
$B\mu$ is fixed to be zero at the GUT scale ($\tan \beta$ is a
prediction rather than a parameter). The two red (solid) lines are the
contours of the Higgs boson mass, 114.1 GeV and 120 GeV. The left side
of the contour $m_h = 114.1$GeV is excluded by current experiments. The
blue (dashed) line denotes the lower bound on
$m_{S}^{2}=-m_{\bar{S}}^{2}$ from the neutral LSP condition
$m_{\tilde{\tau}_1}/m_{\tilde{\chi}}\geq 1$. As we can see, the $D$-term
contribution is large enough to make the stau heavier than the
neutralino, although relatively large ${\rm{tan}}\beta(\gsim 20)$ is
predicted by the condition $B\mu=0$ at the GUT scale. The green (dotted)
lines are the contours of ${\rm BR}(b \ra s \gamma)\times 10^{4} = 2.0$
and $2.5$, respectively. The left side of the contour ${\rm BR}(b \ra s
\gamma)\times 10^{4} = 2.0$ is currently excluded. The orange
(dot-dashed) line is the contour of the selectron mass $m_{\tilde{e}_R}
= 99$GeV, and the inner region surrounded by this line is excluded by
the LEP experiment.  The thin black lines are the contours for $\tan
\beta = 20,22,24,26$, respectively. The black shaded region in the upper
left is where the EWSB does not occur or tachyonic scalars arise. From
Fig. {\ref{fig:Dterm}}, we can see that a parameter region consistent
with phenomenological bounds exist for $M_{1/2} \gsim 350$ GeV and
$m_{S}^2 = -m_{\barr{S}}^2 \gsim (500\GEV)^{2}$. There is still such a
region even if the lower bound on the Higgs boson mass is pushed up to
120 GeV.

Before we close this section, we present some comments.  
Even within the MSSM, we can obtain a SUSY spectrum consistent 
with the Higgs boson mass bound and the requirement of neutralino LSP 
by allowing 
non-zero soft scalar masses for the Higgs fields at the GUT scale.
Such a situation can also be easily realized in gaugino-mediation models
by allowing the Higgs multiplets to propagate in the bulk~\cite{gauginomsb}.
Such a setting may provide a simple solution for generating the $\mu$
term with the correct size by the Giudice-Masiero mechanism~\cite{G-M}.
In this case however, the $B\mu$ term is generally expected to be
non-zero at the boundary and to have an independent phase from the 
$\mu$ term. To solve the SUSY CP problem, this requires an accidental
phase cancellation between the $\mu$ and $B\mu$ terms with
${\cal{O}}(1\%)$.

\section{Conclusions and Discussions}
\setcounter{equation}{0}
Models with the noscale boundary condition naturally solve the
SUSY FCNC problem and possibly also the SUSY CP problem.
Unfortunately, the minimal noscale model was shown to be
not consistent with phenomenological bounds, mainly due to 
the lower bound on the Higgs boson mass and the cosmological 
requirement that the charged particle is not the LSP.
In this paper, we investigate the minimal extension of the 
MSSM with a gauged U(1)$_{B-L}$ symmetry, especially the
SU(5)$\times$U(1)$_{5}$ unification-inspired model, and consider whether 
the noscale boundary condition at the GUT scale is consistent with 
phenomenological constraints or not.

First, we consider the case
with the extended GUT relations given in Eqs.~(\ref{GUT1}) and (\ref{GUT2}),
which are quite natural if the MSSM gauge groups and the U(1)$_{5}$ are
unified into a single group, such as SO(10).
In this case, we find that the particle spectrum is almost the 
same as the minimal noscale model and that there exists almost no
parameter region consistent with the experimental and cosmological 
constraints. 
 
Next, we relax the extended GUT relations and assume that 
the U(1)$_{5}$ is not unified into a single group.
In this case, it is very natural that the gaugino mass for the 
U(1)$_{5}$ is different from those of the MSSM gauge groups, which 
are assumed to be universal, by a factor of order one.
As a result, we find that the stau is heavy enough not
to be the LSP when the gaugino of the U(1)$_{5}$ is 
somewhat heavier than those of the MSSM gauge groups,
and that a wide parameter region is consistent with all the 
constraints. This may imply a SU(5) unification, rather than a SO(10), in
models with the noscale boundary condition.

Finally, we consider the case in which the $S$ and $\bar{S}$ fields,
which are the Higgs fields to break the U(1)$_{5}$ spontaneously at an
intermediate scale,
have non-vanishing soft scalar masses at the GUT scale.
This does not introduce any dangerous flavor-violating interaction,
but provides a large $D$-term contribution which easily 
solves the charged LSP problem, even if we impose the extended GUT
relations given in Eqs.~(\ref{GUT1}) and (\ref{GUT2}).
In this case, we can also impose the condition $B\mu=0$ at the GUT scale 
consistently with the EWSB in a wide parameter region, 
and hence this case is free also from
the SUSY CP problem.

\section*{Acknowledgments}

We are grateful to Tsutomu Yanagida for his suggestion of this 
topic, careful reading of this
manuscript, and continuous encouragements. M.F. and
K.S. are supported by the Japan Society for the Promotion of Science.

\appendix
\section{RGEs for the Yukawa couplings and the soft SUSY breaking terms}
In this appendix, we show the list of the RGEs for 
the Yukawa couplings and soft SUSY breaking terms
for the SU(5)$\times$U(1)$_{5}$ model.
Here, we include the Dirac neutrino Yukawa couplings and their effects 
to the soft SUSY breaking terms for completeness,
which we have neglected in the numerical analyses. 
As for the kinetic term mixings between the two U(1) gauge multiplets,
we neglect them and only show the terms induced from the diagonal parts.
When the extended GUT relation given in Eq.~(\ref{ExGUT}) is
imposed, one can easily include all the mixing effects at one-loop level 
by working in the diagonal basis denoted by Eq.~(\ref{diagonalbasis}).
Even when that relation is not imposed, one can
include the dominant mixing effects on the soft scalar mass terms
by replacing the appropriate terms with the terms given in
Eq.~(\ref{RGEWM}) as we have done in this work.
Our conventions for soft breaking terms except the gaugino masses are given by
\begin{eqnarray}
V_{{\rm{soft}}}=&&\tilde{\bar{u}}A_{u}\tilde{Q}H_{u}-\tilde{\bar{d}}A_{d}
\tilde{Q}H_{d}-\tilde{\bar{e}}A_{e}\tilde{L}H_{d}+
B\mu H_{u} H_{d}+
\tilde{\bar{N}}A_{\nu}\tilde{L}H_{u}+A_{1}\tilde{X}\tilde{S}\tilde{\bar{S}}
+\frac{1}{2}\tilde{S}\tilde{\bar{N}}A_{2}\tilde{\bar{N}}\nonumber\\
&&+\tilde{Q}^{\dag}m_{Q}^{2}\tilde{Q}+\tilde{L}^{\dag}m_{L}^{2}\tilde{L}
+\tilde{\bar{u}}m_{u}^{2}\tilde{\bar{u}}^{\dag}
+\tilde{\bar{d}}m_{d}^{2}\tilde{\bar{d}}^{\dag}
+\tilde{\bar{e}}m_{e}^{2}\tilde{\bar{e}}^{\dag}
+m_{H_{u}}^{2}H_{u}^{*}H_{u}+m_{H_{d}}^{2}H_{d}^{*}H_{d}\nonumber\\
&&+\tilde{\bar{N}}m_{\bar{N}}^{2}\tilde{\bar{N}}^{\dag}
+m_{X}^{2}\tilde{X}^{*}\tilde{X}+m_{S}^{2}\tilde{S}^{*}\tilde{S}
+m_{\bar{S}}^{2}\tilde{\bar{S}}^{*}\tilde{\bar{S}}\;.
\end{eqnarray}
As for the gaugino masses, see Eq.~(\ref{potential}).
\subsection{RGEs for the Yukawa couplings and the $\mu$ term}
\begin{eqnarray}
\frac{d}{dt}(y_{\nu})_{ij}=\frac{1}{16\pi^2}
\left[\begin{array}{lll}
\left\{-3 g_{2}^{2}-\displaystyle{\frac{3}{5}}g_{1}^{2}-\frac{19}{10}g_{5}^{2}
+3{\rm{Tr}}(y_{u}^{\dag}y_{u})+{\rm{Tr}}(y_{\nu}y_{\nu}^{\dag})
\right\}(y_{\nu})_{ij}\\
\\
+3(y_{\nu}y_{\nu}^{\dag}y_{\nu})_{ij}+(y_{\nu}y_{e}^{\dag}y_{e})_{ij}
+(\lambda_{2}\lambda_{2}^{\dag}y_{\nu})_{ij}
\end{array}
\right]
\end{eqnarray}
\begin{eqnarray}
\frac{d}{dt}(y_{e})_{ij}=\frac{1}{16\pi^2}\left[
\begin{array}{lll}
\left\{-3 g_{2}^{2}-\displaystyle{\frac{9}{5}}g_{1}^{2}-\frac{7}{10}g_{5}^{2}
+3{\rm{Tr}}(y_{d}^{\dag}y_{d})+{\rm{Tr}}(y_{e}y_{e}^{\dag})
\right\}(y_{e})_{ij}\\
\\
+3(y_{e}y_{e}^{\dag}y_{e})_{ij}+(y_{e}y_{\nu}^{\dag}y_{\nu})_{ij}
\end{array}
\right]
\end{eqnarray}
\begin{eqnarray}
\frac{d}{dt}(y_{u})_{ij}=\frac{1}{16\pi^2}\left[
\begin{array}{lll}
\left\{-\displaystyle{\frac{13}{15}}g_{1}^{2}-3g_{2}^{2}-\frac{16}{3}g_{3}^{2}
-\frac{3}{10}g_{5}^2
+3{\rm{Tr}}(y_{u}^{\dag}y_{u})+{\rm{Tr}}(y_{\nu}^{\dag}y_{\nu})
\right\}(y_{u})_{ij}\\\\
+3{\rm{Tr}}(y_{u}y_{u}^{\dag}y_{u})_{ij}+(y_{u}y_{d}^{\dag}y_{d})_{ij}
\end{array}
\right]
\end{eqnarray}
\begin{eqnarray}
\frac{d}{dt}(y_{d})_{ij}=\frac{1}{16\pi^2}\left[
\begin{array}{lll}
\left\{-\displaystyle{\frac{7}{15}}g_{1}^{2}-3g_{2}^{2}
-\frac{16}{3}g_{3}^{2}-\frac{7}{10}g_{5}^{2}+3{\rm{Tr}}(y_{d}^{\dag}y_{d})
+{\rm{Tr}}(y_{e}^{\dag}y_{e})
\right\}(y_{d})_{ij}\\
\\
+3(y_{d}y_{d}^{\dag}y_{d})_{ij}+(y_{d}y_{u}^{\dag}y_{u})_{ij}
\end{array}\right]
\end{eqnarray}
\begin{eqnarray}
\frac{d}{dt}(\lambda_{2})_{ij}=\frac{1}{16\pi^2}\left[
\begin{array}{lll}
\left\{
-\displaystyle{\frac{15}{2}}g_{5}^{2}+{\rm{Tr}}
(\lambda_{2}^{\dag}\lambda_{2})+\lambda_{1}^{\dag}\lambda_{1}
\right\}(\lambda_{2})_{ij}\\
\\
+2(\lambda_{2}\lambda_{2}^{\dag}\lambda_{2})_{ij}+2 (y_{\nu}y_{\nu}^{\dag}\lambda_{2})_{ij}+2 (y_{\nu}y_{\nu}^{\dag}\lambda_{2})_{ji}
\end{array}
\right]
\end{eqnarray}
\begin{eqnarray}
\frac{d}{dt}\lambda_{1}=\frac{1}{16\pi^2}\left[
\left\{
-10g_{5}^{2}+{\rm{Tr}}(\lambda_{2}^{\dag}\lambda_{2})
\right\}\lambda_{1}
+3 (\lambda_{1}^{\dag}\lambda_{1})\lambda_{1}
\right]
\end{eqnarray}
\begin{equation}
\frac{d}{dt}\mu=\frac{1}{16\pi^{2}}\mu\left[
{\rm{Tr}}\left[
3y_{u}^{\dag}y_{u}+3y_{d}^{\dag}y_{d}+y_{e}^{\dag}y_{e}+y_{\nu}^{\dag}y_{\nu}
\right]
-3g_{2}^{2}-\frac{3}{5}g_{1}^{2}-\frac{2}{5}g_{5}^{2}
\right]
\end{equation}
\subsection{RGEs for the soft SUSY breaking terms}
\begin{equation}
S\equiv m_{H_{u}}^{2}-m_{H_{d}}^2+{\rm{Tr}}\left[
m_{Q}^{2}+m_{d}^{2}-2 m_{u}^{2}-m_{L}^{2}+m_{e}^{2}
\right]
\end{equation}
\begin{equation}
S_{5}\equiv 4m_{H_{u}}^{2}-4m_{H_{d}}^{2}+10m_{S}^{2}-10m_{\bar{S}}^{2}
+{\rm{Tr}}\left[
-6m_{Q}^{2}-3 m_{u}^{2}+9 m_{d}^{2}+6m_{L}^{2}-m_{e}^{2}-5 m_{\bar{N}}^{2}
\right]
\end{equation}
\begin{eqnarray}
\frac{d}{dt}(m_{e}^{2})_{ij}=\frac{1}{16\pi^2}\left[
\begin{array}{lll}
2 (m_{e}^{2}y_{e}y_{e}^{\dag}+y_{e}y_{e}^{\dag}m_{e}^{2})_{ij}+
4(y_{e}m_{L}^{2}y_{e}^{\dag}+m_{H_{d}}^{2}y_{e}y_{e}^{\dag}
+A_{e}A_{e}^{\dag})_{ij}\\
\\
+\left\{
-\displaystyle{\frac{24}{5}}g_{1}^{2} |M_{1}|^{2}-\frac{1}{5}g_{5}^{2}
|M_{5}|^{2}+\frac{6}{5}g_{1}^{2}S-\frac{1}{20}g_{5}^{2}S_{5}
\right\}\delta_{ij}\end{array}
\right]
\end{eqnarray}
\begin{eqnarray}
\frac{d}{dt}(m_{L}^{2})_{ij}=\frac{1}{16\pi^{2}}\left[
\begin{array}{lll}
(m_{L}^{2}y_{e}^{\dag}y_{e}+y_{e}^{\dag}y_{e}m_{L}^{2})_{ij}
+(m_{L}^{2}y_{\nu}^{\dag}y_{\nu}+y_{\nu}^{\dag}y_{\nu}m_{L}^{2})_{ij}\\\\
+2 (y_{e}^{\dag}m_{e}^{2}y_{e}+m_{H_{d}}^{2}y_{e}^{\dag}y_{e}+
A_{e}^{\dag}A_{e})_{ij}
+2 (y_{\nu}^{\dag}m_{\bar{N}}^{2}y_{\nu}+m_{H_{u}}^{2}y_{\nu}^{\dag}y_{\nu}
+A_{\nu}^{\dag}A_{\nu})_{ij}\\\\
+\left\{
-6g_{2}|M_{2}|^{2}-\displaystyle{\frac{6}{5}}g_{1}^{2}|M_{1}|^{2}
-\frac{9}{5}g_{5}^{2}|M_{5}|^{2}-\frac{3}{5}g_{1}^{2}S+\frac{3}{20}
g_{5}^{2}S_{5}\right\}\delta_{ij}
\end{array}
\right]
\nonumber\\
\nonumber\\
\end{eqnarray}
\begin{eqnarray}
\frac{d}{dt}(m_{\bar{N}}^{2})_{ij}=\frac{1}{16\pi^2}
\left[\begin{array}{lll}
2(m_{\bar{N}}^{2}y_{\nu}^{\dag}y_{\nu}+y_{\nu}y_{\nu}^{\dag}
m_{\bar{N}}^{2})_{ij}
+4(y_{\nu}m_{L}^{2}y_{\nu}^{\dag}+m_{H_{u}}^{2}y_{\nu}y_{\nu}^{\dag}+A_{\nu}A_{\nu}^{\dag})_{ij}\\
\\
+(m_{\bar{N}}^{2}\lambda_{2}\lambda_{2}^{\dag}+\lambda_{2}\lambda_{2}^{\dag}
m_{\bar{N}}^{2})_{ij}
+2(\lambda_{2}m_{\bar{N}}^{2}\lambda_{2}^{\dag}+m_{S}^{2}\lambda_{2}
\lambda_{2}^{\dag}+A_{2}A_{2}^{\dag})_{ij}\\
\\
+\left\{
-5g_{5}^{2}|M_{5}|^{2}-\displaystyle{\frac{1}{4}}g_{5}^{2}S_{5}
\right\}\delta_{ij}
\end{array}
\right]
\end{eqnarray}
\begin{eqnarray}
\frac{d}{dt}(m_{Q}^{2})_{ij}=\frac{1}{16\pi^{2}}\left[
\begin{array}{lll}
(m_{Q}^{2}y_{u}^{\dag}y_{u}+y_{u}^{\dag}y_{u}m_{Q}^{2})_{ij}+
(m_{Q}^{2}y_{d}^{\dag}y_{d}+y_{d}^{\dag}y_{d}m_{Q}^{2})_{ij}\\\\
+2(y_{u}^{\dag}m_{u}^{2}y_{u}+y_{u}^{\dag}y_{u}m_{H_{u}}^{2}+A_{u}^{\dag}A_{u})_{ij}
+2(y_{d}^{\dag}m_{d}^{2}y_{d}+y_{d}^{\dag}y_{d}m_{H_{d}}^{2}+A_{d}^{\dag}A_{d})_{ij}\\\\
+\left\{
-\displaystyle{\frac{2}{15}}g_{1}^{2}|M_{1}|^{2}-6g_{2}^{2}|M_{2}|^{2}
-\frac{32}{3}g_{3}^{2}|M_{3}|^{2}-\frac{1}{5}g_{5}^{2}|M_{5}|^{2}
+\frac{1}{5}g_{1}^{2}S-\frac{1}{20}g_{5}^{2}S_{5}
\right\}\delta_{ij}
\end{array}
\right]\nonumber\\
\nonumber\\
\end{eqnarray}
\begin{eqnarray}
\frac{d}{dt}(m_{d}^{2})_{ij}=\frac{1}{16\pi^2}\left[
\begin{array}{lll}
2(m_{d}^{2}y_{d}y_{d}^{\dag}+y_{d}y_{d}^{\dag}m_{d}^{2})_{ij}
+4(y_{d}m_{Q}^{2}y_{d}^{\dag}+m_{H_{d}}^{2}y_{d}y_{d}^{\dag}+A_{d}
A_{d}^{\dag})_{ij}\\\\
+\left\{
-\displaystyle{\frac{8}{15}}g_{1}^{2}|M_{1}|^{2}-\frac{32}{3}g_{3}^{2}
|M_{3}|^{2}-\frac{9}{5}g_{5}^{2}|M_{5}|^{2}+\frac{2}{5}g_{1}^{2}S+\frac{3}{20}
g_{5}^{2}S_{5}
\right\}\delta_{ij}
\end{array}
\right]
\end{eqnarray}
\begin{eqnarray}
\frac{d}{dt}(m_{u}^{2})_{ij}=\frac{1}{16\pi^{2}}
\left[\begin{array}{lll}
2(m_{u}^{2}y_{u}y_{u}^{\dag}+y_{u}y_{u}^{\dag}m_{u}^{2})_{ij}
+4(y_{u}m_{Q}^{2}y_{u}^{\dag}+m_{H_{u}}^{2}y_{u}y_{u}^{\dag}+A_{u}
A_{u}^{\dag})_{ij}\\\\
+\left\{
-\displaystyle{\frac{32}{15}}g_{1}^{2}|M_{1}|^{2}-\frac{32}{3}g_{3}^{2}|M_{3}|^{2}-\frac{1}{5}g_{5}^{2}|M_{5}|^{2}-\frac{4}{5}g_{1}^{2}S-\frac{1}{20}
g_{5}^{2}S_{5}
\right\}\delta_{ij}
\end{array}
\right]
\end{eqnarray}
\begin{eqnarray}
\frac{d}{dt}m_{H_{d}}^{2}=\frac{1}{16\pi^{2}}
\left[
\begin{array}{lll}
6{\rm{Tr}}\left[m_{H_{d}}^{2}y_{d}^{\dag}y_{d}+y_{d}m_{Q}^{2}y_{d}^{\dag}
+y_{d}^{\dag}m_{d}^{2}y_{d}+A_{d}^{\dag}A_{d}\right]\\\\
+2{\rm{Tr}}\left[
m_{H_{d}}^{2}y_{e}^{\dag}y_{e}+y_{e}m_{L}^{2}y_{e}^{\dag}
+y_{e}^{\dag}m_{e}^{2}y_{e}+A_{e}^{\dag}A_{e}
\right]\\\\
+\left\{
-6g_{2}^{2}|M_{2}|^{2}-\displaystyle{\frac{6}{5}}g_{1}^{2}|M_{1}|^{2}
-\frac{4}{5}g_{5}^{2}|M_{5}|^{2}-\frac{3}{5}g_{1}^{2}S-\frac{1}{10}
g_{5}^{2}S_{5}
\right\}
\end{array}
\right]
\end{eqnarray}
\begin{eqnarray}
\frac{d}{dt}m_{H_{u}}^{2}=\frac{1}{16\pi^{2}}
\left[\begin{array}{lll}
6{\rm{Tr}}\left[m_{H_{u}}^{2}y_{u}^{\dag}y_{u}+y_{u}m_{Q}^{2}y_{u}^{\dag}
+y_{u}^{\dag}m_{u}^{2}y_{u}+A_{u}^{\dag}A_{u}
\right]\\\\
+2{\rm{Tr}}\left[m_{H_{u}}^{2}y_{\nu}y_{\nu}^{\dag}+
y_{\nu}m_{L}^{2}y_{\nu}^{\dag}+y_{\nu}^{\dag}m_{\bar{N}}^{2}y_{\nu}
+A_{\nu}^{\dag}A_{\nu}
\right]\\\\
+\left\{
-6g_{2}^{2}|M_{2}|^{2}-\displaystyle{\frac{6}{5}}g_{1}^{2}|M_{1}|^{2}
-\frac{4}{5}g_{5}^{2}|M_{5}|^{2}+\frac{3}{5}g_{1}^{2}S+\frac{1}{10}
g_{5}^{2}S_{5}\right\}
\end{array}
\right]
\end{eqnarray}
\begin{eqnarray}
\frac{d}{dt}m_{S}^{2}=\frac{1}{16\pi^{2}}\left[
\begin{array}{lll}
2 m_{S}^{2}{\rm{Tr}}(\lambda_{2}^{\dag}\lambda_{2})
+2{\rm{Tr}}\left[
\lambda_{2}^{\dag}m_{\bar{N}}^{2}\lambda_{2}+\lambda_{2}m_{\bar{N}}^{2}
\lambda_{2}^{\dag}+A_{2}^{\dag}A_{2}\right]\\\\
+2(m_{S}^{2}+m_{X}^{2}+m_{\bar{S}}^{2})\lambda_{1}^{\dag}\lambda_{1}+
2 A_{1}^{\dag}A_{1}\\\\
+\left\{
-20g_{5}^{2}|M_{5}|^{2}+\frac{1}{2}g_{5}^{2}S_{5}
\right\}
\end{array}
\right]
\end{eqnarray}
\begin{eqnarray}
\frac{d}{dt}m_{\bar{S}}^{2}=\frac{1}{16\pi^{2}}\left[
\begin{array}{lll}
2(m_{S}^{2}+m_{X}^{2}+m_{\bar{S}}^{2})\lambda_{1}^{\dag}\lambda_{1}
+2 A_{1}^{\dag}A_{1}\\\\
+\left\{
-20g_{5}^{2}|M_{5}|^{2}-\displaystyle{\frac{1}{2}}g_{5}^{2}S_{5}
\right\}
\end{array}
\right]
\end{eqnarray}
\begin{eqnarray}
\frac{d}{dt}m_{X}^{2}=\frac{1}{16\pi^2}\left[
2(m_{X}^{2}+m_{S}^{2}+m_{\bar{S}}^{2})\lambda_{1}^{\dag}\lambda_{1}+2A_{1}^{\dag}A_{1}
\right]
\end{eqnarray}
\begin{eqnarray}
\frac{d}{dt}(A_{e})_{ij}=\frac{1}{16\pi^2}\left[
\begin{array}{lll}
\left[3{\rm{Tr}}(y_{d}^{\dag}y_{d})+{\rm{Tr}}(y_{e}^{\dag}y_{e})
\right](A_{e})_{ij}
+2\left[3{\rm{Tr}}(y_{d}^{\dag}A_{d})+{\rm{Tr}}(y_{e}^{\dag}A_{e})
\right](y_{e})_{ij}\\\\
+4(y_{e}y_{e}^{\dag}A_{e})_{ij}+5(A_{e}y_{e}^{\dag}y_{e})_{ij}
+2(y_{e}y_{\nu}^{\dag}A_{\nu})_{ij}+(A_{e}y_{\nu}^{\dag}y_{\nu})_{ij}
\\\\
+\left\{-3g_{2}^{2}-\displaystyle{\frac{9}{5}}g_{1}^{2}-\frac{7}{10}g_{5}^{2}
\right\}(A_{e})_{ij}
+2\left\{3g_{2}^{2} M_{2}+ \displaystyle{\frac{9}{5}}g_{1}^{2}M_{1}+\frac{7}{10}
g_{5}^{2}M_{5}
\right\}(y_{e})_{ij}
\end{array}
\right]
\nonumber\\
\nonumber\\
\end{eqnarray}
\begin{eqnarray}
\frac{d}{dt}(A_{\nu})_{ij}=\frac{1}{16\pi^{2}}\left[
\begin{array}{lll}
\left[3{\rm{Tr}}(y_{u}^{\dag}y_{u})+{\rm{Tr}}(y_{\nu}^{\dag}y_{\nu})
\right](A_{\nu})_{ij}
+2\left[3{\rm{Tr}}(y_{u}^{\dag}A_{u})+{\rm{Tr}}(y_{\nu}^{\dag}A_{\nu})
\right](y_{\nu})_{ij}\\\\
+4(y_{\nu}y_{\nu}^{\dag}A_{\nu})_{ij}+5(A_{\nu}y_{\nu}^{\dag}y_{\nu})_{ij}
+2(y_{\nu}y_{e}^{\dag}A_{e})_{ij}+(A_{\nu}y_{e}^{\dag}y_{e})_{ij}
\\\\
+(\lambda_{2}\lambda_{2}^{\dag}A_{\nu})_{ij}+2(A_{2}\lambda_{2}^{\dag}y_{\nu})_{ij}+\left\{
-3 g_{2}^{2}-\displaystyle{\frac{3}{5}}g_{1}^{2}-\frac{19}{10}g_{5}^{2}
\right\}(A_{\nu})_{ij}\\\\
+2\left\{3g_{2}^{2}M_{2}+\displaystyle{\frac{3}{5}}g_{1}^{2}M_{1}
+\frac{19}{10}g_{5}^{2}M_{5}
\right\}(y_{\nu})_{ij}
\end{array}
\right]
\end{eqnarray}
\begin{eqnarray}
\frac{d}{dt}(A_{u})_{ij}=\frac{1}{16\pi^{2}}
\left[\begin{array}{lll}
3{\rm{Tr}}(y_{u}y_{u}^{\dag})(A_{u})_{ij}+5(A_{u}y_{u}^{\dag}y_{u})_{ij}
+(A_{u}y_{d}^{\dag}y_{d})_{ij}\\\\
+6{\rm{Tr}}(A_{u}y_{u}^{\dag})(y_{u})_{ij}+4(y_{u}y_{u}^{\dag}A_{u})_{ij}
+2(y_{u}y_{d}^{\dag}A_{d})_{ij}\\\\
+\left\{-\displaystyle{\frac{16}{3}}g_{3}^{2}-3g_{2}^{2}-\frac{13}{15}
g_{1}^{2}-\frac{3}{10}g_{5}^{2}
\right\}(A_{u})_{ij}\\\\
+2\left\{\displaystyle{\frac{16}{3}}g_{3}^{2}M_{3}+3g_{2}^{2}M_{2}+\frac{13}{15}g_{1}^{2}M_{1}
+\frac{3}{10}g_{5}^{2}M_{5}
\right\}(y_{u})_{ij}
\end{array}
\right]
\end{eqnarray}
\begin{eqnarray}
\frac{d}{dt}(A_{d})_{ij}=\frac{1}{16\pi^{2}}
\left[\begin{array}{lll}
{\rm{Tr}}\left[3y_{d}^{\dag}y_{d}
+y_{e}y_{e}^{\dag} \right](A_{d})_{ij}
+5(A_{d}y_{d}^{\dag}y_{d})_{ij}+(A_{d}y_{u}^{\dag}y_{u})_{ij}\\\\
+{\rm{Tr}}\left[6 A_{d}y_{d}^{\dag}+2 A_{e}y_{e}^{\dag}
\right](y_{d})_{ij}+4(y_{d}y_{d}^{\dag}A_{d})_{ij}
+2(y_{d}y_{u}^{\dag}A_{u})_{ij}\\\\
+\left\{-\displaystyle{
\frac{16}{3}}g_{3}^{2}-3g_{2}^{2}-\frac{7}{15}g_{1}^{2}-\frac{7}{10}
g_{5}^{2}
\right\}(A_{d})_{ij}\\\\
+2\left\{\displaystyle{\frac{16}{3}}g_{3}^{2}M_{3}+3g_{2}^{2}M_{2}+\frac{7}{15}
g_{1}^{2}M_{1}+\frac{7}{10}g_{5}^{2}M_{5}
\right\}(y_{d})_{ij}
\end{array}
\right]
\end{eqnarray}
\begin{equation}
\frac{d}{dt}(A_{2})_{ij}=\frac{1}{16\pi^{2}}\left[
\begin{array}{lll}
{\rm{Tr}}(\lambda_{2}^{\dag}\lambda_{2})(A_{2})_{ij}+3(\lambda_{2}
\lambda_{2}^{\dag}A_{2})_{ij}+3(A_{2}\lambda_{2}^{\dag}\lambda_{2})_{ij}\\\\
+2{\rm{Tr}}(\lambda_{2}^{\dag}A_{2})(\lambda_{2})_{ij}+
(\lambda_{1}^{\dag}\lambda_{1})(A_{2})_{ij}+2(\lambda_{1}^{\dag}A_{1})(\lambda_{2})_{ij}\\\\
+2(y_{\nu}y_{\nu}^{\dag}A_{2})_{ij}+2(y_{\nu}y_{\nu}^{\dag}A_{2})_{ji}
+4(A_{\nu}y_{\nu}^{\dag}\lambda_{2})_{ij}+4(A_{\nu}y_{\nu}^{\dag}\lambda_{2})_{ji}\\\\
+\left\{-\displaystyle{\frac{15}{2}}g_{5}^{2}
\right\}(A_{2})_{ij}+
2\left\{\displaystyle{\frac{2}{15}}g_{5}^{2}M_{5}\right\}(\lambda_{2})_{ij}
\end{array}
\right]
\end{equation}
\begin{equation}
\frac{d}{dt}A_{1}=\frac{1}{16\pi^{2}}
\left[\begin{array}{lll}
3(\lambda_{1}^{\dag}\lambda_{1})A_{1}+6(\lambda_{1}\lambda_{1}^{\dag}A_{1})
+{\rm{Tr}}(\lambda_{2}^{\dag}\lambda_{2})A_{1}
+2{\rm{Tr}}(\lambda_{2}^{\dag}A_{2})\lambda_{1}\\\\
+\left\{-10g_{5}^{2}
\right\}A_{1}
+2\left\{10g_{5}^{2}M_{5}
\right\}\lambda_{1}
\end{array}
\right]
\end{equation}
\begin{eqnarray}
\frac{d}{dt}B\mu=\frac{1}{16\pi^{2}}
\left[\begin{array}{lll}
B\mu\left\{
{\rm{Tr}}\left[
3y_{u}^{\dag}y_{u}+3y_{d}^{\dag}y_{d}+y_{e}^{\dag}y_{e}+
y_{\nu}^{\dag}y_{\nu}
\right]-3g_{2}^{2}-\displaystyle{\frac{3}{5}}g_{1}^{2}-\frac{2}{5}g_{5}^{2}
\right\}\\\\
+\mu\left\{{\rm{Tr}}\left[
6y_{u}^{\dag}A_{u}+6y_{d}^{\dag}A_{d}+2y_{e}^{\dag}A_{e}+2y_{\nu}^{\dag}
A_{\nu}
\right]+6g_{2}^{2}M_{2}+\displaystyle{\frac{6}{5}}g_{1}^{2}M_{1}
+\frac{4}{5}g_{5}^{2}M_{5}
\right\}
\end{array}
\right]\nonumber\\
\end{eqnarray}

%
%
\newcommand{\Journal}[4]{{\sl #1} {\bf #2} {(#3)} {#4}}
\newcommand{\PL}{\sl Phys. Lett.}
\newcommand{\PR}{\sl Phys. Rev.}
\newcommand{\PRL}{\sl Phys. Rev. Lett.}
\newcommand{\NP}{\sl Nucl. Phys.}
\newcommand{\ZP}{\sl Z. Phys.}
\newcommand{\PTP}{\sl Prog. Theor. Phys.}
\newcommand{\NC}{\sl Nuovo Cimento}
\newcommand{\MPL}{\sl Mod. Phys. Lett.}
\newcommand{\PRep}{\sl Phys. Rept.}

\newpage
\small
\newpage
\end{document}